\documentclass[12pt,preprint]{aastex}
\doublespace

\usepackage{epsfig}
\usepackage{natbib}
\usepackage{wasysym}
\usepackage{subfigure}
\usepackage{verbatim}
\usepackage{enumerate}

\slugcomment{ }

\shorttitle{Solar Magnetized ``Tornadoes''}
\shortauthors{Su et al.}

\begin{document}

\title{Solar Magnetized ``Tornadoes'': Relation to Filaments}

\author{Yang Su\altaffilmark{1}}
\email{yang.su@uni-graz.at}

\author{Tongjiang Wang\altaffilmark{2,3}}
\author{Astrid Veronig\altaffilmark{1}}
\author{Manuela Temmer\altaffilmark{1}}
\and
\author{Weiqun Gan\altaffilmark{4}}


\altaffiltext{1}{IGAM-Kanzelh\"{o}he Observatory, Institute of Physics, University of Graz, Universit\"{a}tsplatz 5, 8010 Graz, Austria.}
\altaffiltext{2}{Department of Physics, the Catholic University of America, Washington, DC 20064, USA.}
\altaffiltext{3}{Solar Physics Laboratory (Code 671), Heliophysics Science Division, NASA Goddard Space Flight Center, Greenbelt, MD 20771, USA.}
\altaffiltext{4}{Key Laboratory of Dark Matter and Space Astronomy, Purple Mountain Observatory, Chinese Academy of Sciences, Nanjing 210008, China.}

\begin{abstract}
Solar magnetized ``tornadoes'', a phenomenon discovered in the solar atmosphere, appear as tornado-like structures in the corona but root in the photosphere. Like other solar phenomena, solar tornadoes are a feature of magnetized plasma and therefore differ distinctly from terrestrial tornadoes. Here we report the first analysis of solar ``tornadoes''\footnote{Two papers which focused on different aspect of solar tornadoes were published in the Astrophysical Journal Letters \citep{li12} and Nature \citep{wed12}, respectively, during the revision of this Letter.}. A detailed case study of two events indicates that they are rotating vertical magnetic structures probably driven by underlying vortex flows in the photosphere. They usually exist as a group and relate to filaments/prominences, another important solar phenomenon whose formation and eruption are still mysteries. Solar tornadoes may play a distinct role in the supply of mass and twists to filaments. These findings could lead to a new explanation to filament formation and eruption. 
\end{abstract}
 

\section{Solar magnetized ``tornadoes''}

The Atmospheric Imaging Assembly \citep[AIA, ][]{lem12} on board the Solar Dynamics Observatory (SDO) observes the Sun in high temporal cadence ($\sim$12 seconds) and spatial resolution ($\sim$0.6 arcsec). This capability enables detailed observation of a new phenomenon in the solar corona, solar magnetized ``tornadoes''. They appear as tornado-like magnetic structures and widely distribute over the solar disk. Similar phenomenon may have been seen (e.g., the movie 13 ``Tornadoes and fountains in a filament on 2 August 2000'' on the page \url{http://soi.stanford.edu/results/SolPhys200/Schrijver/TRACEsolarphysicsCD.html}) in the observations from the Transition Region and Coronal Explorer \citep[TRACE, ][]{han99} but never been reported in scientific journals. Here we present the first study on solar magnetized ``tornadoes'' and their role in the formation and eruption of filaments/prominences, another solar phenomenon. 

During June 22-23 2011, a group of solar tornadoes developed on the solar northwest limb (Fig. 1 and online movies). The ``tornado'' funnels appear as dark, cone-shaped column structures in the AIA Fe IX 171 {\AA} ($\sim$0.63 MK) passband (Fig.1a-c), connecting the surface and the prominence above seen in AIA He II 304 Å ($\sim$0.05 MK) and Ca II K3 (3934 {\AA}) lines (Fig.1d, f). The largest tornado (located at [683, 628] in Fig.1c) has a height of about 60 arcsec ($\sim$44,260 km) with a width of only $\sim$2 arcsec ($\sim$1,480 km) at bottom and $\sim$10 arcsec ($\sim$7,380 km) at the top. The widths in AIA 304 {\AA}, H-alpha and Ca II K3 are about 2-3 times larger, suggesting that the core of tornadoes may have a lower temperature than that of the surrounding plasma.

AIA 171 {\AA} movies (M1) reveal apparent rotating motions in the evolution of these structures. On June 22 2011, several dark vertical structures appeared on the surface (Fig.1a), and grew to higher altitude. The motion of the attached branches indicates rotation of the whole structure. The overlaying prominence, which connected these tubes, became more prominent on June 23 2011 (see movies in AIA 304 {\AA}). We analyzed the dynamics of solar tornadoes using the time-distance diagrams (see Fig. 2), known as stack plots. The dark sinusoidal features in these plots suggest that the dark vertical threads in solar tornadoes were rotating. Four regions (a, b, c and d) were selected to show the detail (the third and fourth rows in Fig. 2). In particular, the features in region c appear to reveal two mutually rotating threads, which cause the upper part of the tornado to become highly twisted as indicated by the converging motion (with a speed of about 1.5 km s$^{-1}$) shown in time-distance plot (top panel of Fig. 2). Sometimes, the dark threads in solar tornadoes also show divergence (second panel of Fig. 2), which may suggest untwisting motions. If interpreting the sinusoidal features seen in time-distance plots as rotating motions, we can estimate the rotating angular speed ($\omega$) and linear speed ($v$) from the period ($T$) and amplitude ($r$) using $\omega=2\pi/T$ and $v=2{\pi}r/T$. The periods are different for different tornadoes, suggesting that they rotated with different (angular) speeds. For the features in the four regions their periods are measured to be about 70, 50, 70, and 30 minutes, and the estimated rotation speeds\footnote{These values are much smaller than the flow speed of up to 95 km s$^{-1}$ along the helical structure in a prominence and associated cavity reported by \cite{li12}, but is consistent with the average horizontal velocity of a solar tornado in the simulation presented by \cite{wed12}.} of the threads are about 8, 6, 7, and 6 km s$^{-1}$.

However, oscillation motions, which have been reported by variety of researches (see \citealt{iso06,ning09,tri09,boc11} and references therein), cannot be ruled out in our events. They may co-exist with the rotational motion and cause the periodic variations seen in Fig. 2 as signature of torsional Alfv\'{e}n wave in the corona \citep{jess09}. 

To investigate whether solar ``tornadoes'' rotate and whether they relate to the magnetized vortexes detected in the solar photosphere and chromosphere \citep{bon08,bon10,ste10,wed09}\footnote{The new paper by \cite{wed12} has pointed out that the relatively small-scale solar tornadoes they found in observation and simulation are indeed the consequences of the vortex flow on the photosphere.}, we studied another group observed during October 18-21 2010 by multiple spacecrafts at different locations (see Fig. 3 , online movie M2 and M3 for an overview of this event). The AIA 171 {\AA} movie and selected frames in Fig. 4 show rotating threads in one of the solar tornadoes (indicated by a white box in Fig. 3) from 12:50 to 13:50 UT on October 19. By tracking the motions of magnetic elements on the solar surface using the DAVE method \citep{sch06}, horizontal flow map is derived from the photospheric magnetograms observed by the Helioseismic and Magnetic Imager \citep[HMI, ][]{schou12} on board the SDO around 13:11 UT (Fig. 4b and c). It reveals vortex flows with velocity of up to a few hundred meters per second in the footpoint region of the tornado. The coincidence in position and timing suggests that the rotational motion in solar tornadoes has a close connection with the vortex flows in the photosphere. 

\section{Solar tornadoes and filaments}

Strikingly, we found that solar ``tornadoes'' are often correlated with filaments, another important phenomenon in solar corona. Filaments are dark channels filled with relatively cool (around 10$^4$ K) and dense plasma suspended in the hot ($\sim$1 MK) and tenuous corona. When they appear above the solar limb, they are recognized as bright structures called prominences. Filaments and prominences may erupt from the solar surface to become coronal mass ejection (CME) and produce solar flares. Therefore, they are important for understanding the nature of solar activities and predicting the space weather at Earth. How filaments form and evolve to eruption is still a big mystery although various models have been proposed (see \citealt{lab10} and \citealt{mac10} for a review). Filaments are believed to be helical magnetic flux ropes \citep{rust94}. The source of dense, cold plasma and twists in the filament, however, is still unclear. 

Spine and barbs are the two structural components of a filament. Spine is the highest, horizontal part of a filament while barbs connect spine and the surface. Barbs are important for understanding filament formation and the plasma flows in filaments. But the nature of barbs remains unknown. Barbs often root in network boundaries among three or more supergranular cells \citep{plo73,lin05} in the photosphere. On the other hand, vortexes are also located in such boundaries of granular or supergranular cells. However, the connection between barbs and vortex has never been built. 

The close relationship between solar ``tornadoes'' and filaments has been shown in the 2011-June-23 event. In the event on October 18-21 2010 (Fig. 3), their relations are revealed more clearly using observations from simultaneous two views with a separation angle of 80 degrees between the Solar Terrestrial Relations Observatory \citep[STEREO-B/EUVI, ][]{wue04} and SDO/AIA (aided with ground-based H-alpha observations). The association with solar tornadoes may provide a new key to understand the formation and eruption of filaments.

Below is what we learned from this event:
1) Barbs appear earlier than the filament spine itself and may act as supporting stands during the evolution of the filament. Only three dark blobs (barbs) were visible in H-alpha on October 18. One day later (October 19), line structures (spine fields) were seen between barbs in H-alpha and 171 {\AA} images (Fig. 3), especially between the two southern barbs. On October 20, the filament with three major barbs was clearly visible in the H-alpha spectral line. 2) The filament barbs, as viewed from side, are the vertical tornado structures and appear in projection as viewed from Earth. Previous studies have found counter-streaming as evidence for vertical magnetic structure in filaments \citep{zir98}. Here we present in Fig. 3 (October 18 and 19) the good agreement in spatial relation between the filament barbs (H-alpha) and the solar ``tornadoes'' (AIA 171 {\AA}). Particularly, Fig. 3 (October 20) shows that the locations of barbs as viewed on-disk have no corresponding horizontal features when viewed from side (STEREO-B), but vertical tornado structures. Best evidence for projection effects is given by the most northern tornado. Its coronal part, as viewed from Earth, maps to the most northern on-disk barb location. 3) Solar tornadoes can erupt with the associated filament. The movies recorded in SDO/AIA 171 (M2) and STEREO-B 195 {\AA} (M3) show the filament formation and eruption. Two ribbons and bright loops are seen on October 22 after the eruption (online movie M2), which are similar to that produced in a solar flare. However, no significant X-ray emission was detected. 

Fig. 5 shows additional examples for the eruption of solar tornadoes and associated filaments. In the first example, the tornadoes started to form not later than October 15 2010, and erupted on October 26 2010. In the second example, a group of large tornadoes were associated with a giant filament across the solar disk, which appeared in SDO field of view on November 07 2011. A partial eruption cleaned out the east part of the filament on November 14 2011. These eruptions show a common feature that a major tornado in the group appears to rise first, together with the overlying filament and cavity. The erupting filament then 'pulls up' other tornadoes. However, a recent study by \cite{reg11} showed that the eruption of a cavity and filament is a two-stage process composed of a slow rise phase and an acceleration phase. Thus it cannot exclude the possibility that the rising tornado is a consequence of filament eruption. In whatever case the continuous rotation of solar tornadoes can effectively build up twists in the filament structure (the slow rising phase), and finally lead to their eruption due to MHD instability.

\section{Summary and Discussion}

Thanks to two space missions, SDO and STEREO, we obtained direct images of solar magnetized ``tornadoes'' from two different angles. The observations show possible connections among different solar features, such as solar tornadoes, vortexes, cavities and filaments. The results are summarized below:

 \begin{enumerate}[1.]
\item Solar ``tornadoes'' are rotating magnetic structures in solar atmosphere. Their lifetime could be from hours to weeks (small scale tornadoes with shorter lifetime may also exist but could be difficult to detect\footnote{\cite{wed12} did find solar tornadoes with lifetime of about 10 minutes.}). They usually exist as a group of tornadoes due to their magnetic connections. It is unknown whether or not isolated tornado exists. Vortex flow is found at the foot of solar tornado as evidence for the rotational motion of magnetic structures. The driver of solar ``tornadoes'' is very likely associated with vortexes at granules and supergranules boundaries in the photosphere. The period is about tens of minutes and the rotation speed about 5-10 km s$^{-1}$. One of the measured tornadoes seems to show acceleration in (angular) speed with time.
\item A group of solar tornadoes exists below filament spine and evolves with it, from formation to eruption (a short-lived event may show no appearance of a filament). Barbs are actually the projections of the tornado funnels as viewed from Earth. They appear before filament spine forms (contains enough cold plasma that absorbs radiation from below to be visible as dark structures). Solar tornadoes can erupt with filament spine. These eruptions generally produce two ribbons and loops, similar to those in a solar flare, but no significant enhancement in GOES soft X-rays. The evolution of filament indicates that cold plasma and twists may be transported from the surface into filament spine through rotating tornado funnels to make the filament visible as dark structure in H-alpha line. 
\end{enumerate}

These observations may suggest a filament model in which magnetized tornadoes (or vortexes in the photospheric networks) play the central role in filament formation and eruption. The vortex flows converge and twist the magnetic tubes which connect different vortexes. As they rotate, cold plasma and twists could be transported into filament spine through tornado funnels. Barbs then appear first on the surface, and spine afterwards. The filament becomes highly twisted and unstable. It may erupt eventually with these underlying tornado funnels. 

This picture needs to be further tested by observations and simulations. The challenges in this interpretation are: 1) The model requires some upflows in the vortexes for the mass supply while most observations \citep{bon08,bon10} and simulations \citep{kit11,kit12,moll11,moll12,she11} show downflows in vortex core (so-called ``sinkhole'') at photosphere height. It is noticed that these research also show signatures for some upflows in the surrounding area of vortex core. Besides, upflow was seen in vortex at chromospheric height \cite{wed09}. There are also reports on upflow at the feet of filament \citep{zir98,cao10}. These results may suggest height-dependent vertical flows in vortexes\footnote{New observations reported by \cite{wed12} provided evidence for upflow in tornado structures with a typical speed of 4 km s$^{-1}$ at chromospheric height.}. 2) The lifetime of solar tornadoes is much longer than the small-scale vortexes related to granules (in the order of several minutes). It is possible that larger vortexes with longer lifetime exist in supergranule scale as observed before \citep{att09,tian10}. What we observed in the corona may relate to some of the ``solar cyclones'', which are also rotating magnetic structures with relatively long lifetime of more than 10 hours \citep{zha11}. 3) The magnetic vortexes could exist anywhere in the photospheric networks while filaments always form above Polarity Inversion Lines \citep[PILs, ][]{zir98}. One explanation for the fact that solar tornadoes prefer to form as a group may be that the emergence of a twisted flux rope \citep{oka10,ber11} below the surface during filament formation produces a group of vortexes (and tornadoes). 

Continuous measurements of Doppler velocity in the ``tornado'' structures from spectrographic instruments that cover for at least several hours are needed for future studies. However, it should be pointed out that rotating motion revealed from Doppler signal may not necessarily be an indicator for the existence of tornado structure. A ``magnetized tornado'' refers to rotating magnetic structures or magnetic funnels while plasma flows along static helical flux ropes could generate equivalent Doppler feature.

The connection with other solar features through magnetic field is the most important characteristic of solar tornado, like other solar phenomena. However, it is unknown whether or not all filaments are associated with solar tornadoes. What we report here may provide an important clue to the future development of models and simulations on filaments and magnetized tornadoes themselves. 

  
\acknowledgments

We thank the referee for the valuable comments. The space observations used in this study were obtained from the Solar Dynamics Observatory (SDO) and the Solar Terrestrial Relations Observatory (STEREO). SDO is a mission for NASA's Living With a Star (LWS) Program. The ground observations were obtained from the National Solar Observatory -GONG, the Observatory of Paris-Meudon in France, and the Kanzelh\"{o}he Observatory (KSO)/University of Graz in Austria.
This work utilizes data obtained by the Global Oscillation Network Group (GONG) Program, managed by the National Solar Observatory, which is operated by AURA, Inc. under a cooperative agreement with the National Science Foundation. The data were acquired by instruments operated by the Big Bear Solar Observatory, High Altitude Observatory, Learmonth Solar Observatory, Udaipur Solar Observatory, Instituto de Astrof\'\i sica de Canarias, and Cerro Tololo Interamerican Observatory. 
This activity has been supported by the European Community Framework Programme 7, High Energy Solar Physics Data in Europe (HESPE), grant agreement no.: 263086. The work of T. W. was supported by NASA grants NNX10AN10G and NNX12AB34G. M. T. greatly acknowledges the Austrian Science Fund (FWF): FWF V195-N16. W. G. acknowledges 2011CB811402 by MSTC.


\clearpage

 
\begin{figure}
\begin{center}
\epsscale{1.}
\plotone{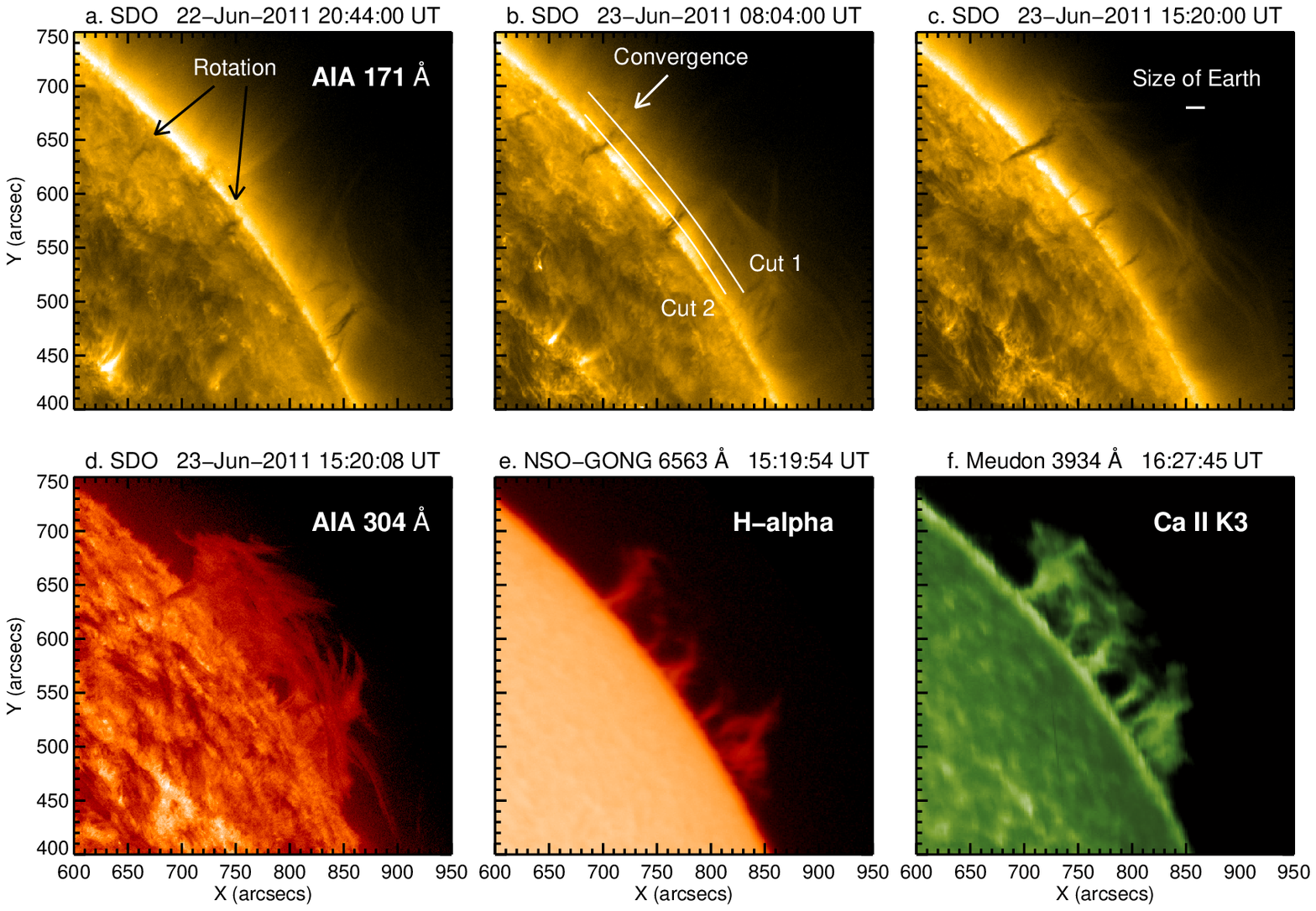}
\caption{Example of a group of solar tornadoes observed on 22-23 June 2011. a-c: selected images in SDO/AIA Fe IX 171 {\AA} ($\sim$0.63 million kelvins, MK). Two lines, ``cut1'' and ``cut2'', are used to obtain the time-distance diagrams shown in Fig. 2. One arc second refers to about 738 km on the Sun. d-f: Images in He II (304 {\AA}, SDO/AIA, $\sim$0.05 MK), H-alpha (6562.8 {\AA}, the National Solar Observatory, NSO-GONG) and Ca II K3 (3934 {\AA}, the Observatory of Paris-Meudon).\label{fig1}}
\end{center}
\end{figure}

\clearpage


\begin{figure}
\epsscale{.80}
\plotone{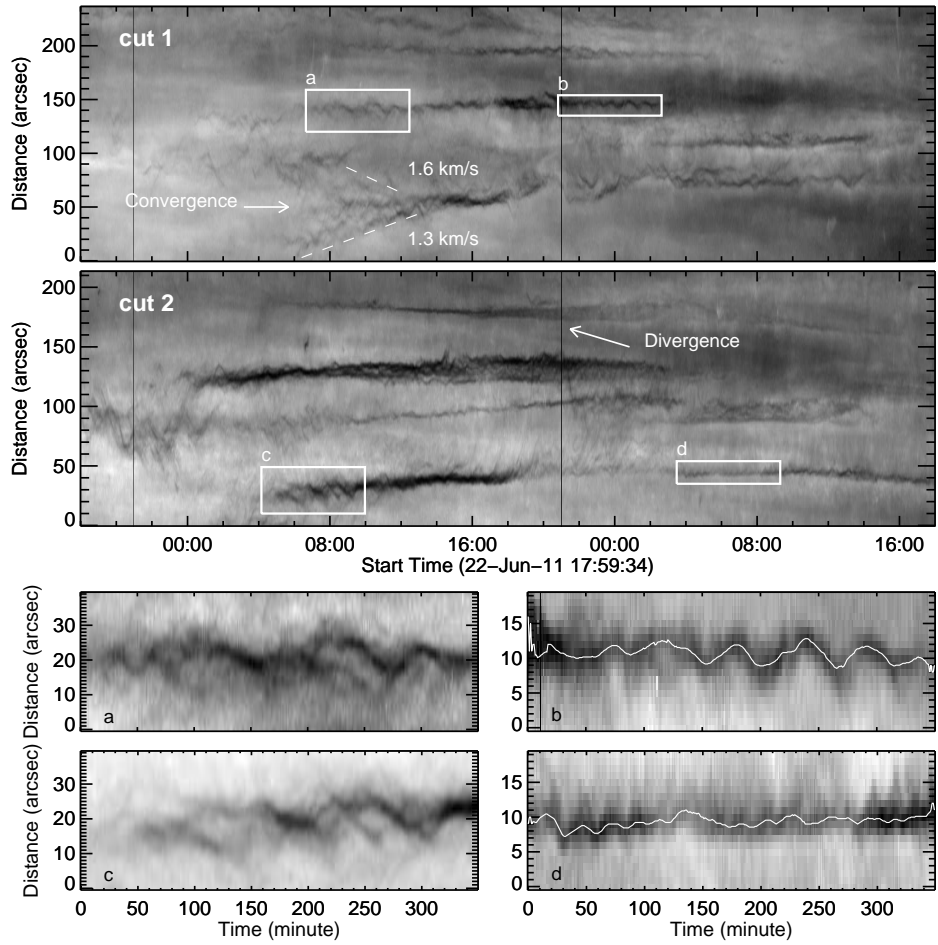}
\caption{Dynamics of solar tornadoes on 22-23 June 2011. First and second panels: Time-distance diagrams (stack plots) along ``cut1'' and ``cut2'' in Fig. 1B (with start point at the north end). Third and fourth panels: Four regions a, b, c and d were selected to show the details of the periodic variations.\label{fig2}}
\end{figure}

\clearpage
 
 \begin{figure}
 \begin{center}
\includegraphics[scale=.80]{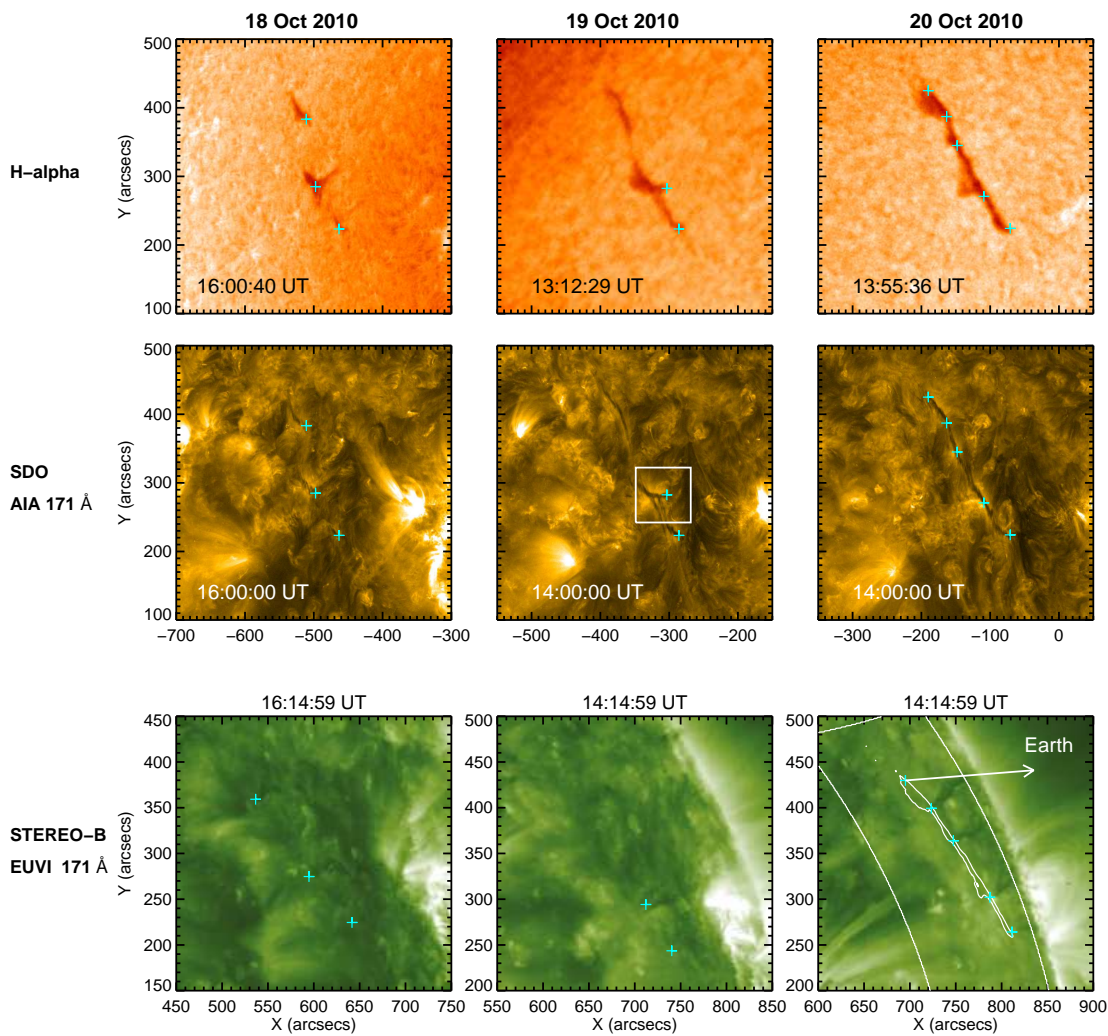}
\caption{Evolution of solar tornadoes and related filament from 18 to 20 October 2010. The images (in rows) are obtained in the H-alpha spectral line (the Big Bear Solar Observatory, BBSO; the Observatory of Paris-Meudon; and the Kanzelh\"{o}he Observatory, KSO), SDO/AIA 171, and STEREO-B 171 {\AA}, respectively. The crosses show reference points for footpoint areas of the dark vertical structures. The white box indicates the region for 171 {\AA} images in Fig. 4. The white contour shows the location and shape of the filament viewed from Earth as would be observed from STEREO-B at the same time. One arc second is about 722 km on the Sun for this date.}
\end{center}
\end{figure}

\clearpage

 \begin{figure}
 \begin{center}
\includegraphics[scale=.70]{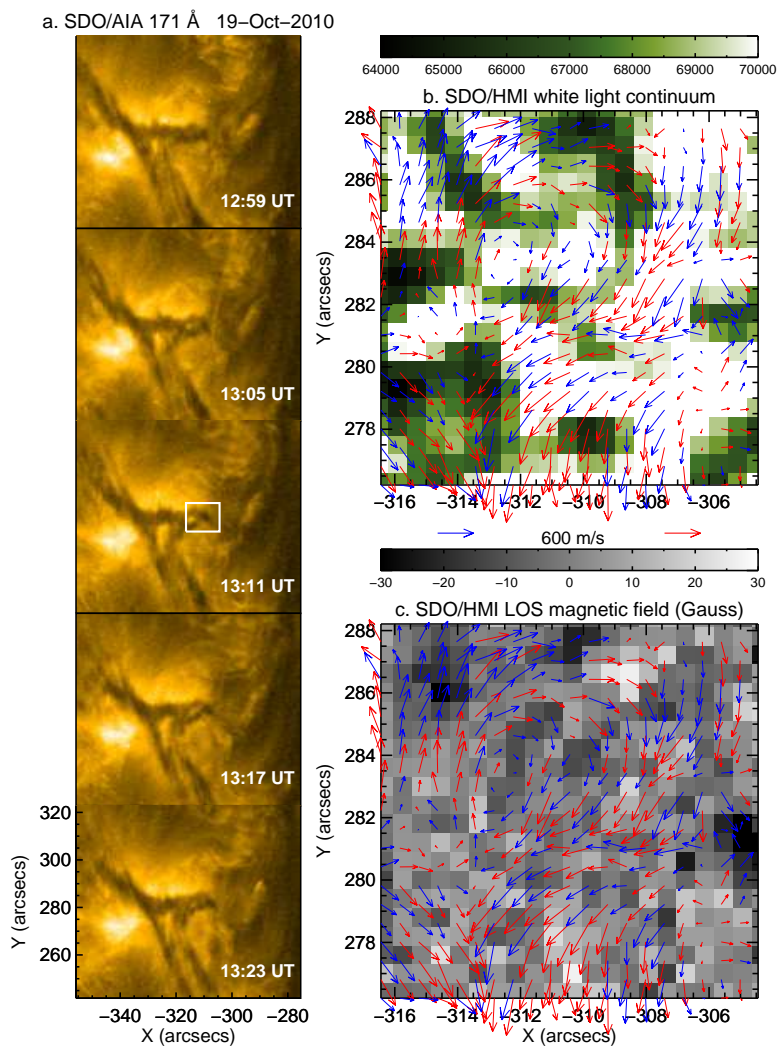}
\caption{Evidence for rotational motion in a solar tornado on October 19 2010. a: Selected AIA 171 {\AA} images from 12:59 to 13:23 UT. The white box in the image at 13:11 UT indicates location of the region shown on the right. b and c: Horizontal velocity field (longest arrow represents 600 m/s) superimposed on SDO/HMI white light continuum image (b) and SDO/HMI LOS magnetic field (c) map. Red arrows stand for positive magnetic field and blue for negative. The velocities are derived from HMI magnetic measurements around 13:11 UT (10 min cadence) and the DAVE method (with a window size of 10 pixels).}
\end{center}
\end{figure}

\clearpage

 \begin{figure}
 \begin{center}
\includegraphics[scale=.80]{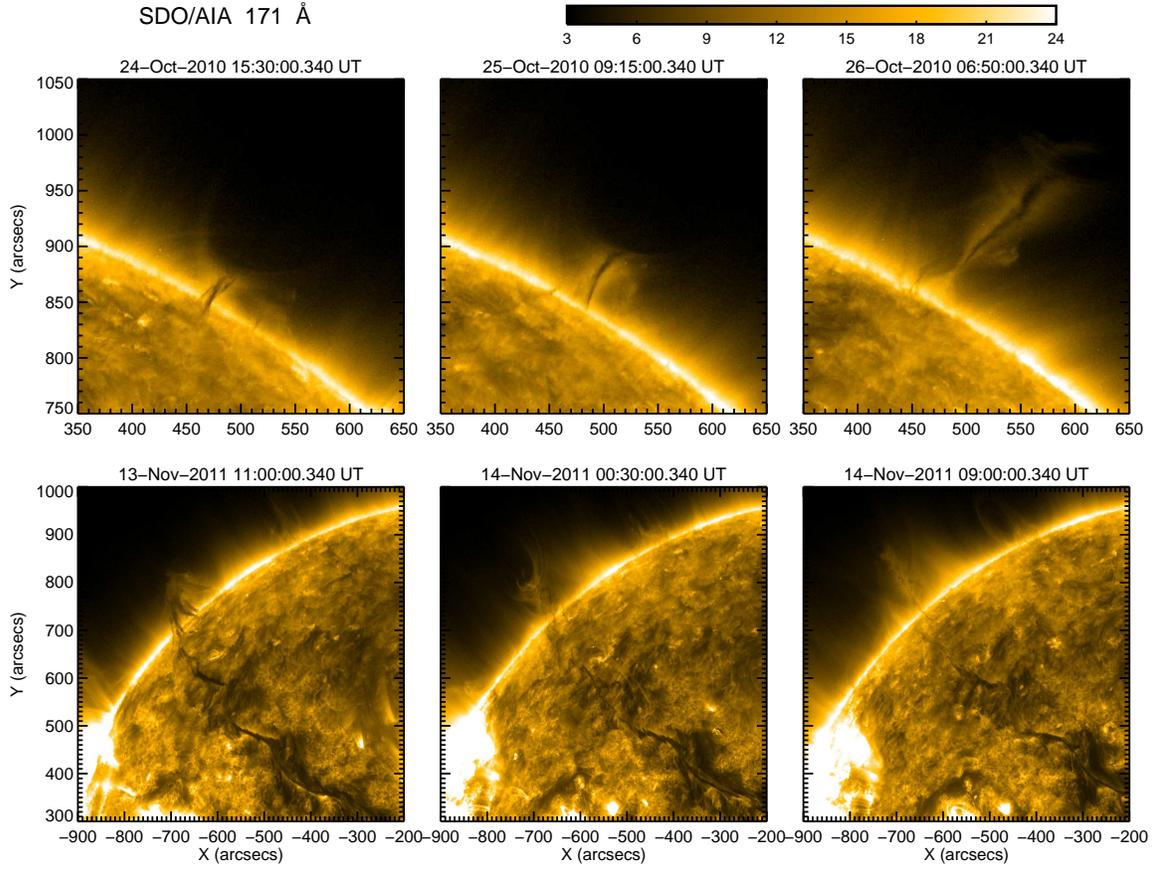}
\caption{Two examples for the eruption of solar tornadoes and associated filaments. The images were taken at SDO/AIA 171 {\AA} on 24-26 October 2010 and 13-14 November 2011, respectively.}
\end{center}
\end{figure}

 
\end{document}